\documentclass[useAMS]{mn2e}
\usepackage{lscape}
\usepackage{rotating}
\usepackage{amssymb}
\usepackage{float}
\usepackage{hyperref}
\usepackage{newtxtext,newtxmath}

\def\msun{\hbox{M$_\odot$}}

\def\t4{\hbox{t$_{\rm 4}$}}

\def\cm3{\hbox{cm$^{-3}$}}

\input psfig.sty
\input epsf.sty
\usepackage{graphicx}
\usepackage{epsfig}
\title[MPs in NGC7252:W3]
{Searching for Multiple Populations in the Integrated Light of the Young and Extremely Massive Clusters in the Merger Remnant NGC~7252}
\author[Bastian et al.] {Nate Bastian$^1$,  Carmela Lardo$^2$, Christopher Usher$^1$, Sebastian Kamann$^1$, S{\o}ren S. Larsen$^3$, \newauthor Ivan Cabrera-Ziri$^4$\thanks{Hubble fellow}, William Chantereau$^1$, Silvia Martocchia$^{1,5}$, Maurizio Salaris$^1$, \newauthor Randa Asa'd$^6$, and Michael Hilker$^5$\\
$^{1}$Astrophysics Research Institute, Liverpool John Moores University, 146 Brownlow Hill, Liverpool L3 5RF, UK\\
$^{2}$Laboratoire d'astrophysique, Ecole Polytechnique F\'ed\'erale de Lausanne (EPFL), Observatoire de Sauverny, CH-1290 Versoix, Switzerland\\
$^{3}$Department of Astrophysics/IMAPP, Radboud University, P.O. Box 9010, 6500 GL Nijmegen, The Netherlands\\
$^{4}$ Harvard-Smithsonian Center for Astrophysics, 60 Garden Street, Cambridge, MA 02138, USA\\
$^{5}$European Southern Observatory, Karl-Schwarzschild-Stra\ss e 2, D-85748 Garching bei M\"unchen, Germany\\
$^{6}$Physics Department, American University of Sharjah, P.O. Box 26666, Sharjah, UAE\\
}
\date{Accepted. Received; in original form}
\pagerange{\pageref{firstpage}--\pageref{lastpage}}
\pubyear{2019}
\begin{document}
\maketitle
\label{firstpage}
\begin{abstract}
Recent work has shown that the properties of multiple populations within massive stellar clusters (i.e., in the extent of their abundance variations as well as the fraction of stars that show the anomalous chemistry) depend on the mass as well as the age of the host cluster.  Such correlations are largely unexpected in current models for the formation of multiple populations and hence provide essential insight into their origin.  Here we extend or previous study into the presence or absence of multiple populations using integrated light spectroscopy of the $\sim600$~Myr, massive ($\sim10^7 - 10^8$~\msun) clusters, W3 and W30, in the galactic merger remnant, NGC 7252.  Due to the extreme mass of both clusters, the expectation is that they should host rather extreme abundance spreads, manifested through, e.g., high mean [Na/Fe] abundances.  However, we do not find evidence for a strong [Na/Fe] enhancement, with the observations being consistent with the solar value.  This suggests that age is playing a key role, or alternatively that multiple populations only manifest below a certain stellar mass, as the integrated light at all ages above $\sim100$~Myr is dominated by stars near or above the main sequence turn-off.

\end{abstract}

\section{Introduction}
\label{sec:intro}

Recent results have demonstrated that the phenomenon of multiple populations (MPs) in massive clusters is not limited to the ancient globular clusters, in the Milky Way or other galaxies, but also occurs in clusters down to an age of at least $\sim2$~Gyr, or a formation redshift of $z\sim0.17$.  This has been found through photometric studies of resolved massive clusters in the Large and Small Magellanic Clouds (LMC/SMC; Niederhofer et al.~2017; Martocchia et al.~2018a,2019, Li \& de Grijs~2019) which primarily trace nitrogen (N) variations between stars, spectroscopic measurements directly probing N-variations (Hollyhead et al.~2017; 2018), also in LMC/SMC clusters, and also through integrated light studies sensitive to Na variations (Bastian et al.~2019).  

In the latter study, we used high and low resolution spectra of a massive ($1.7\times10^7$~\msun), intermediate age ($\sim3$~Gyr) cluster in the merger remnant, NGC~1316 (cluster G114), along with stellar population models tailored for the cluster, with and without the characteristic abundance anomalies (see Bastian \& Lardo~2018 for a recent review) to show that the cluster is strongly enhanced in Na and likely Al.  This is expected if MPs are present.  Using the same techniques, we also studied the integrated spectrum of NGC~1978, a $\sim2.2$~Gyr cluster in the LMC with a mass of $\sim3\times10^5$~\msun\ and did not find any evidence for significant Na enhancement within the cluster.  This is in agreement with resolved abundances analysis of individual stars within the cluster  (Mucciarelli et al.~2008), although this cluster does display MPs in the form of small nitrogen variations (Martocchia et al.~2018a,b).  One of the clear differences between NGC~1978 and G114 is their mass, with G114 being nearly $100$ times more massive than NGC~1978.  Hence, it may be that G114 hosts more extreme MPs than NGC~1978 due to its much higher mass, as the range of abundance variations and the fraction of anomalous stars within clusters both correlate with (present day) cluster mass (Carretta et al.~2010; Schiavon et al.~2013; Milone et al.~2017; Martocchia et al.~2019).

However, cluster age also appears to play an important role in determining when, and to what extent, MPs are present within a cluster (Martocchia et al.~2018a, 2019; Li \& de Grijs~2019).  These authors found a clear correlation between the photometric dispersion of the RGB in LMC/SMC clusters in UV/blue filters (chosen for their sensitivity to the N abundance of stars along the RGB) and cluster age, with older clusters having significantly larger spreads than younger clusters.  Additionally, to date, no clusters below an age of $\sim2$~Gyr have been found to host abundance spreads\footnote{While younger clusters do display features in their colour-magnitude diagrams that are not well reproduced by a simple stellar population (e.g., extended main sequence turn-offs and split main sequences) these appear to be caused by a different mechanism, likely stellar rotation and not chemical variations inherited from their initial composition.  See Bastian \& Lardo~(2018) for a recent review.}, although whether they do not exist within the clusters or are just too small to see with the current errors, remains an open question.

Here, we extend our analysis to the extremely massive clusters, W3 ($8\times10^7$~\msun - Maraston et al.~2004) and W30 ($1.6\times10^7$~\msun - Bastian et al.~2006), in the galactic merger remnant NGC~7252.  These clusters have ages of $600\pm100$~Myr and solar metallicity (Schweizer \& Seitzer~1998; Cabrera-Ziri et al.~2016a).

This paper is organised as follows: in \S~\ref{sec:obs} we introduce the observations and models used, while in \S~\ref{sec:results} we present our main results and discuss their implications.  We present our conclusions in \S~\ref{sec:conclusions}.

\section{Observations and Models}
\label{sec:obs}

\subsection{Observations}


We use the UVES spectrum of NGC~7252:W3 that has been discussed in detail in Maraston et al.~(2004 - Program ID: 67.D-0205(A)), with a total exposure time of 7.65 hours.  Briefly, the spectrum covered the wavelength range $3500-6600$\AA\ with a resolution of 2~km/s at 5200\AA.   For W30, we use the UVES spectrum presented in Bastian et al.~(2006 - Program ID: 073.D-0305(B)).   The wavelength range covered was $4200 - 6200$\AA\ at a resolution of $5$~km/s at $5200$\AA\ ($R\sim60,000$) with a total on-target exposure time of 25.1 hours.

The large radial velocities of W3 and W30, with respect to the Milky Way, means that the clusters' spectral features are redshifted out of the range where Galactic ISM absorption is a potential problem. The S/N (per \AA) is $\sim150$ in the NaD region of the spectra for W3 and $\sim90$ for W30.  Both clusters have low extinction values - obtained by comparing their UV and optical colours with simple stellar population models (Bastian et al.~2013), implying low ISM absorption by the host galaxy.  Additionally, any ISM absorption would only serve to increase the strength of the NaD lines, whereas (as discussed later in the paper) the measured NaD strength does not show indications of enhanced values.

Additionally, VLT/MUSE observations of NGC 7252 have been obtained under program ID 099.B-0281 (PI: Privon). The galaxy was observed with a 2x2 mosaic, with each of the four pointings being exposed for approximately 5500s. We downloaded the reduced data cube of the full mosaic from the ESO archive. It has an estimated seeing of 1.1".  

We extracted the spectrum of W3 from the MUSE datacube using a circular aperture with a radius of 4 spaxels (0.8"), centered on the visually defined cluster position. A background galaxy spectrum was created by averaging all spaxels in an annulus with an inner radius of 8 spaxels (1.6") and an outer radius of 14 spaxels (2.8"). It was subtracted from the spectrum of W3 prior to any analysis. We verified that the results did not depend significantly on the extracted radii that were chosen.  The extracted spectrum has S/N (per \AA) of $90$ in the NaD region.  The extracted spectrum covers the wavelength region from $4800-9300$\AA\ at a constant FWHM of 2.5\AA\, meaning that the resolution varies from $R\sim1700$ in the blue to $R\sim3500$ in the red.

Finally, we used archival data of the cluster NGC7252:W3 taken with the Xshooter spectrograph on the Very Large Telescope under ESO programme number 087.B- 0758(C). The observations took place between 24 July 2011 and 18 August 2011, using the 0.8~arcsec $\times$ 11~arcsec slit for the UVB arm and the 0.9~arsec $\times$ 11~arcsec slit for the VIS arm, resulting to a cumulative exposure time of 5.85 hours. During the observations, the seeing ranged between 0.79 and 1.5~arcsec while the airmass ranged between 1.119 and 1.436.

The data reduction was done with the standard Xshooter Reflex pipeline (version 2.9.0) where each individual exposure was reduced independently in stare mode to produce a1d spectra. This reduction consisted in subtraction of bias and dark frames, flat fielding, order extraction and rectification and flux and wavelength calibration. We combined the individual 1d exposures using {\sc scombine} in {\sc iraf}, using a median as the method. We also allowed for $\pm3$~sigma clipping to minimise the effect of cosmic rays that were not successfully corrected by the pipeline in the earlier steps.  The S/N (per \AA) of the extracted spectrum of W3 is 42 near NaD.

\begin{table} 
  \begin{tabular}
    {llllll} Galaxy & Cluster & Age & Mass & $\sigma$ & v$_{\rm r}$\\
    & & (Gyr) & (\msun) & (km/s) & (km/s) \\
    \hline
 NGC~7252 & W3 & 0.6 & $8\times10^7$ & $45$ & 4822\\
 NGC~7252 & W30 & 0.5 & $1.6\times10^7$ & $27.5$ & 4614\\
 \hline
  \end{tabular}
\caption{Properties of the clusters studied in the present work.  Both clusters have solar metallicity.  References for the values are given in the text.}
\label{tab:properties}
\end{table}

\subsection{Stellar Population Models}
\label{sec:models}

\begin{figure}
\centering
\includegraphics[width=8cm]{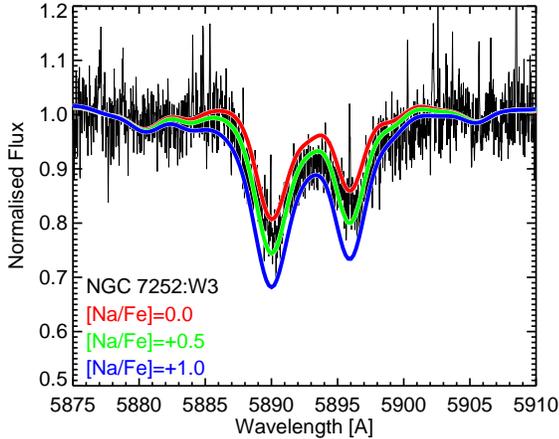}
\caption{The NaD line of W3, along with integrated stellar population models with different levels of multiple populations, from solar abundance ratios (red) to intermediate (green) and extreme (blue).  The models adopt an age of 600~Myr and [Fe/H]$=0.0$.  The effect of the first dredge-up on the surface abundance of stars past the main sequence turn-off has not been included.  When viewing in black and white, in this and all figures with models, the legend lists the models in order of their appearance in the cores of absorption lines, from top to bottom.}
\label{fig:nad}
\end{figure}

The models used in the current work were developed in the same way as those used in Bastian et al.~(2019) and we refer the interested reader to that work for more details.  In summary, we developed integrated spectra of stellar populations at a given age, [Fe/H] and chemical composition.  We explored chemical compositions made to mimic standard element abundances (i.e. "primordial" or 1P stars of "field-like abundances") as well as that of multiple populations observed in GCs.  For the latter, we generate an "intermediate" population with [N/Fe] = [Na/Fe] = [Al/Fe] = 0.5, [C/Fe] = [O/Fe] = -0.09, and [Mg/Fe] = -0.07 and an "extreme" population with  [N/Fe] = [Na/Fe] = [Al/Fe] = 1.0, [C/Fe] = [O/Fe] = -0.70, [Mg/Fe] = -0.44.   We show the best fitting (see \S~\ref{sec:best_fit_w3}) models (600~Myr, solar metallicity) for the three cases of MPs (no MPs, intermediate and extreme) and the observed NaD profile of W3 in Fig.~\ref{fig:nad}.

The models were initially computed for R=200,000 and were then convolved with the velocity dispersion measured for each of the clusters, $\sigma_{\rm W3}=45$~km/s (Maraston et al.~2004) and $\sigma_{\rm W30} = 27.5$~km/s (Bastian et al.~2006).

In \S~\ref{sec:dredge_up} we will discuss further additions to the modelling due to the age of the cluster, namely the inclusion of the effects on the surface abundances due to the first dredge-up.

\subsection{Selecting the best fit model for W3}
\label{sec:best_fit_w3}

W3 has been the subject of a number of studies which have attempted to estimate its age, metallicity and mass through integrated light meaurements (Miller et al.~1996; Schweizer \& Seitzer~1998; Maraston et al.~2004; Cabrera-Ziri et al.~2016a).  Each of these studies has used different SSP models to estimate the cluster's parameters, although each has suggested an age of $500-600$~Myr and a metallicity near solar.  Around this age/metallicity, there are systematic differences between SSP models which can be traced to different underlying assumptions (such as the library of stellar spectra and the theoretical stellar isochrones - e.g., Conroy \& Gunn~2010).  Hence, the cluster age/metallicity estimates found in the literature may not be consistent with our modelling of the integrated spectrum, 
because the theoretical isochrones and spectra we employ in this analysis may differ from those used in previous works.

As discussed in \S~\ref{sec:models} we have developed our own set of SSP models that, for a given age and [Fe/H], allow us to change the abundance of individual elements in order to mimic the abundance patterns characteristic of MPs in GCs.  We will refer to these models as the {\sc Chrusher} models.

\begin{figure}
\centering
\includegraphics[width=8cm]{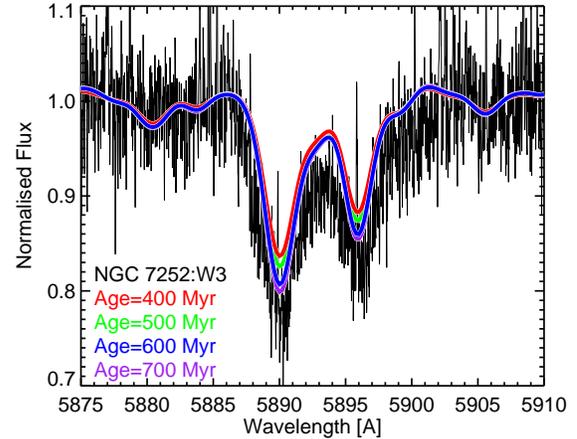}

\caption{The NaD spectral features for NGC~7252:W3 in the UVES spectrum along with the computed synthetic spectra for four different ages (all with [Fe/H]=0 and no MPs).  The effect of the first dredge-up has not been included here.}
\label{fig:w3_age}
\end{figure}

\begin{figure}
\centering
\includegraphics[width=8cm]{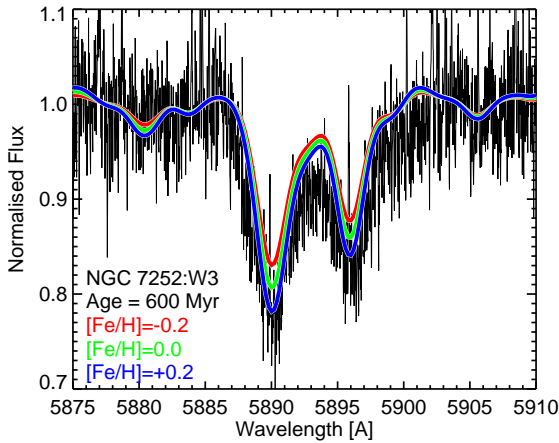}

\caption{The NaD spectral features for NGC~7252:W3 along with the computed synthetic spectra for three different metallicities (all with an age of 600~Myr and no MPs).  As in Fig.~\ref{fig:w3_age} the effect of the first dredge-up has not been included here.}
\label{fig:w3_z}
\end{figure}

Using the {\sc Chrusher} models, we show the effect of small changes in the age, at fixed metallicity, in the integrated light NaD feature in Fig.~\ref{fig:w3_age}.   Similarly, the effect of metallicity, at fixed age, is shown in Fig.~\ref{fig:w3_z}.  As can be seen, both parameters can have a relatively strong effect on the NaD feature and the adoption of a wrong model may lead us to erroneous conclusions.

In order to find the best fitting model, we have used the {\sc ULySS} spectral fitting tool (Koleva et al.~2008).  We have adopted the Pegase HR simple stellar population models from Le Borgne et al. (2004) based on the Elodie~3.1 library of stellar spectra (Prugniel \& Soubiran~2001).  All Pegase HR models are computed assuming a Kroupa~(2001) IMF and Padova 1994 (Bertelli et al.~1994))  evolutionary tracks providing synthetic SSPs with ages between $1-20,000$ Myr and metallicities between $-2.3$ and $0.69$~dex.

\begin{figure}
\centering
\includegraphics[width=9cm]{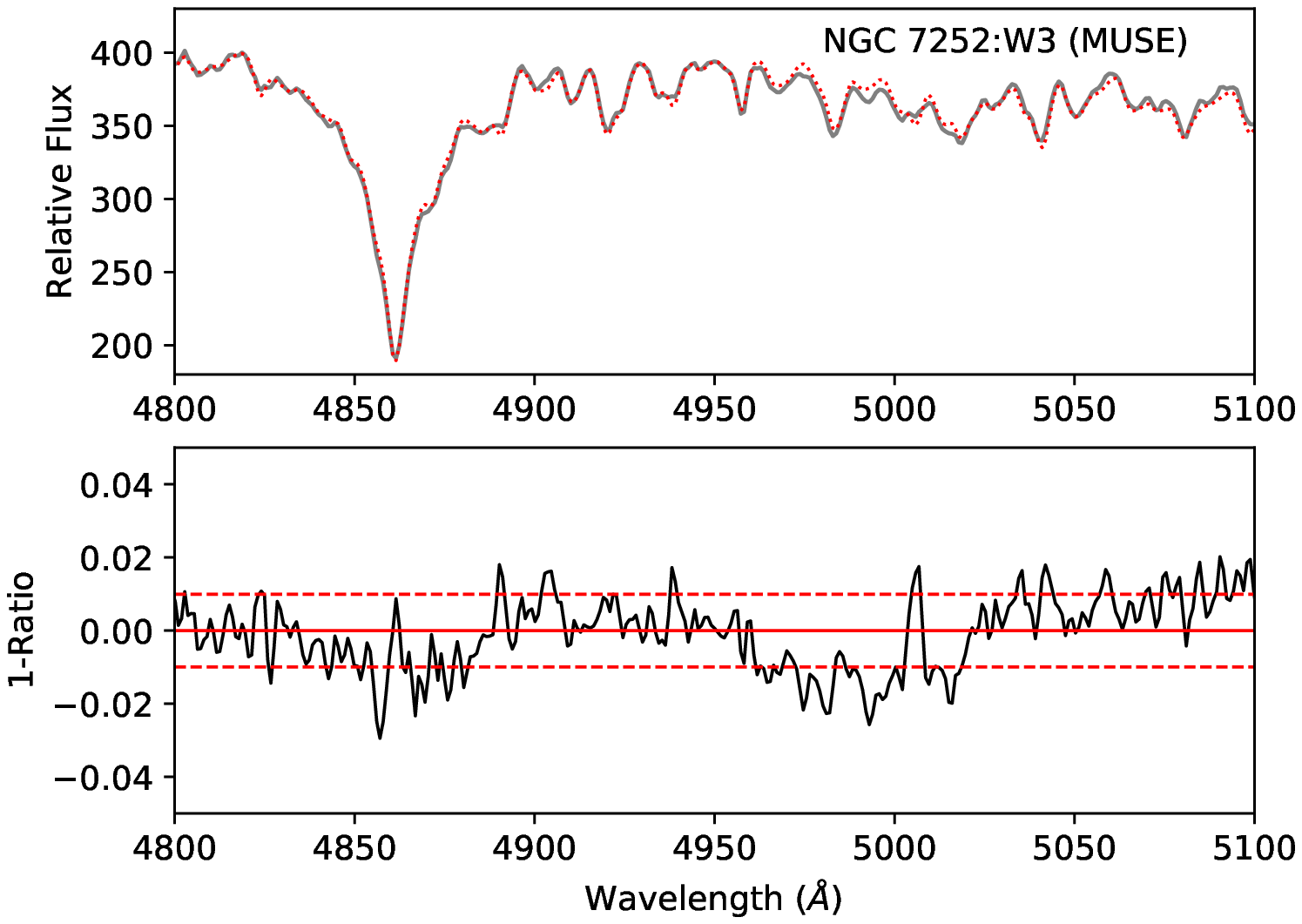}
\includegraphics[width=9cm]{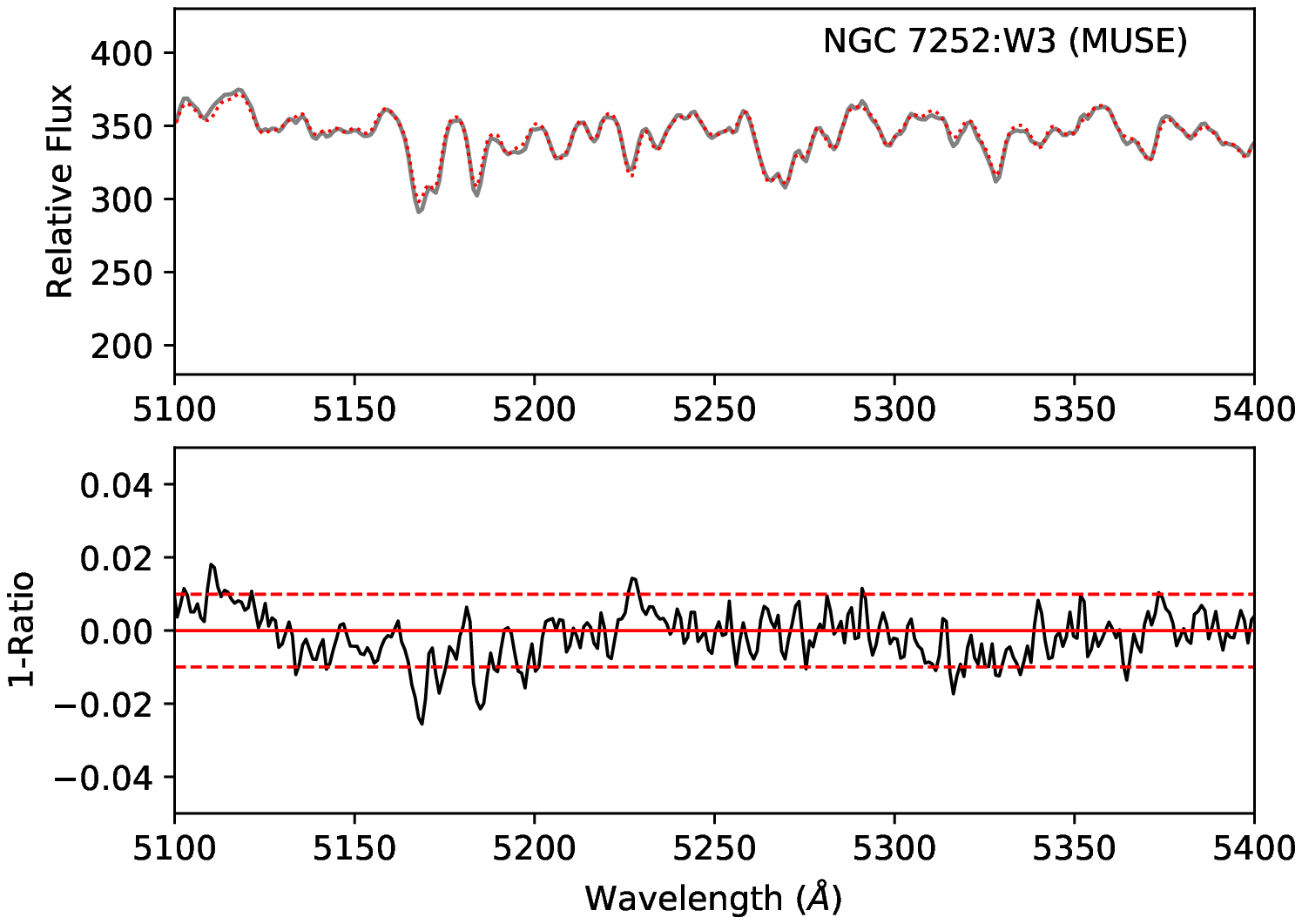}
\includegraphics[width=9cm]{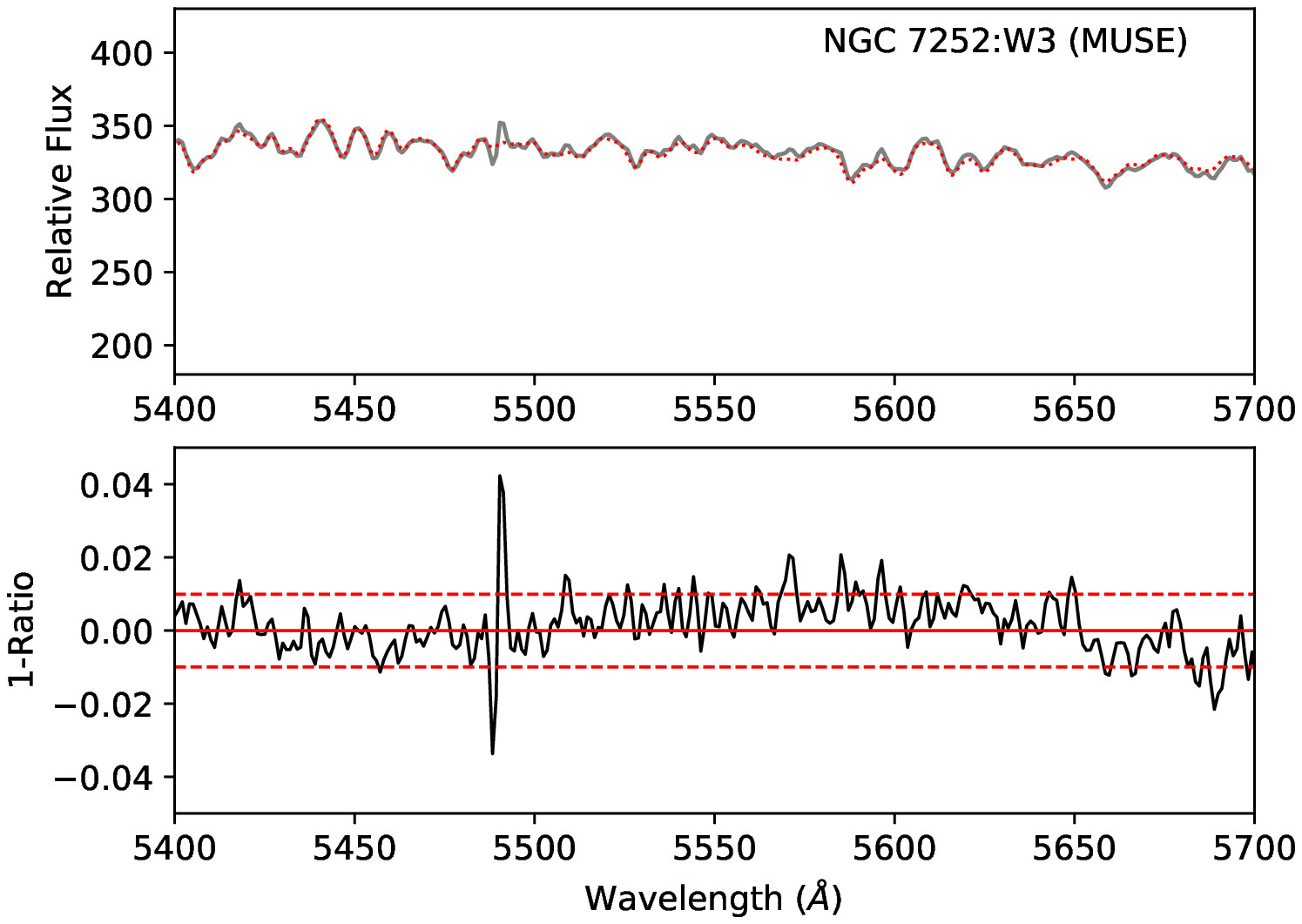}

\caption{Three portions of the MUSE spectrum of W3 (shown in grey), along with the best fit from {\sc ULySS} (shown in red - dotted line).  In the bottom panels we show the one minus the ratio between the observations and the best fitting model.  The solid red line in the bottom panels shows the zero level while the dashed horizontal lines represent the 1$\sigma$ deviation.}
\label{fig:w3_muse}
\end{figure}

We begin by fitting the observed Xshooter and MUSE spectra of W3. The NaD lines were masked in the process as they are also strongly affected by the presence of MPs, if present.   The best fit parameters are shown in Table~\ref{tab:comparison}.  We obtain consistent results for both spectra, with an age of $\sim518$~Myr and solar metallicity.  In Fig.~\ref{fig:w3_muse} we show part of the MUSE spectrum along with the best fit model as well as the residuals of the fit and we note the overall excellent fit to the observations. 

We then ran the code on our model spectra ({\sc Chrusher}), for different ages ($400-700$~Myr), metallicities ([Fe/H]=$-0.2, 0, +0.2$), with and without MPs.  The results are given in Table~\ref{tab:comparison}.  We see a systematic offset, where the ages derived by {\sc ULySS} (using the Pegase HR SSPs) are systematically younger than the ages of the {\sc Chrusher} models.  We find the {\sc Chrusher} model with the closest derived properties to the observed W3 spectrum has an age of 600~Myr with solar metallicitiy (with or without MPs).

The implication of these tests is that the closest {\sc Chrusher} model spectra to the observed W3 spectra has an age of $600$~Myr and solar metallicity.  We will adopt this set of parameters as our default model for W3 throughout the paper, and use it to estimate the effect on the spectrum when the chemistry of multiple populations is added.  The age difference found between {\sc ULySS} and {\sc Chrusher} can be used as a rough estimate of the zero point error associated with the choice of SSP models. 

\begin{table*} 
  \begin{tabular}
    {llllll} Observation & Age$_{Chrusher}$ & [Fe/H]$_{Chrusher}$ &Age$_{ULySS}$  & [Fe/H]$_{ULySS}$ \\
    &  Myr & dex & Myr  \\
    \hline
W3 (Xshooter)  	       & $\hdots$ & $\hdots$ &   505   &  0.00 \\ 
W3 (MUSE)	       & $\hdots$ & $\hdots$ &   532   & -0.03 \\ 
		       &        &	 &	   &       \\
Models                 &        &        &         &  \\  
\hline
No MPs		       & 400    &  0.0   &   381   &  0.01\\  
No MPs		       & 500    &  0.0   &   483   &  0.02\\  
No MPs		       & 500    & -0.2   &   477   & -0.14\\  
No MPs		       & 500    & +0.2   &   434   &  0.23\\  
MPs intermediate       & 500    &  0.0   &   481   &  0.00\\  
MPs extreme	       & 500    &  0.0   &   474   &  0.00\\  
No MPs		       & 600    &  0.0   &   567   &  0.05\\  
No MPs		       & 600    & -0.2   &   548   & -0.11\\  
No MPs                 & 600    & +0.2   &   491   &  0.30\\  
MPs intermediate       & 600    &  0.0   &   556   &  0.05\\  
MPs extreme	       & 600    &  0.0   &   577   &  0.01\\  
No MPs		       & 700    &  0.0   &   661   &  0.06\\  
\hline

  \end{tabular}
\caption{Comparison of the best fitting templates for the observations and {\sc Chrusher} models.} 
\label{tab:comparison}
\end{table*}

\begin{figure}
\centering
\includegraphics[width=9cm]{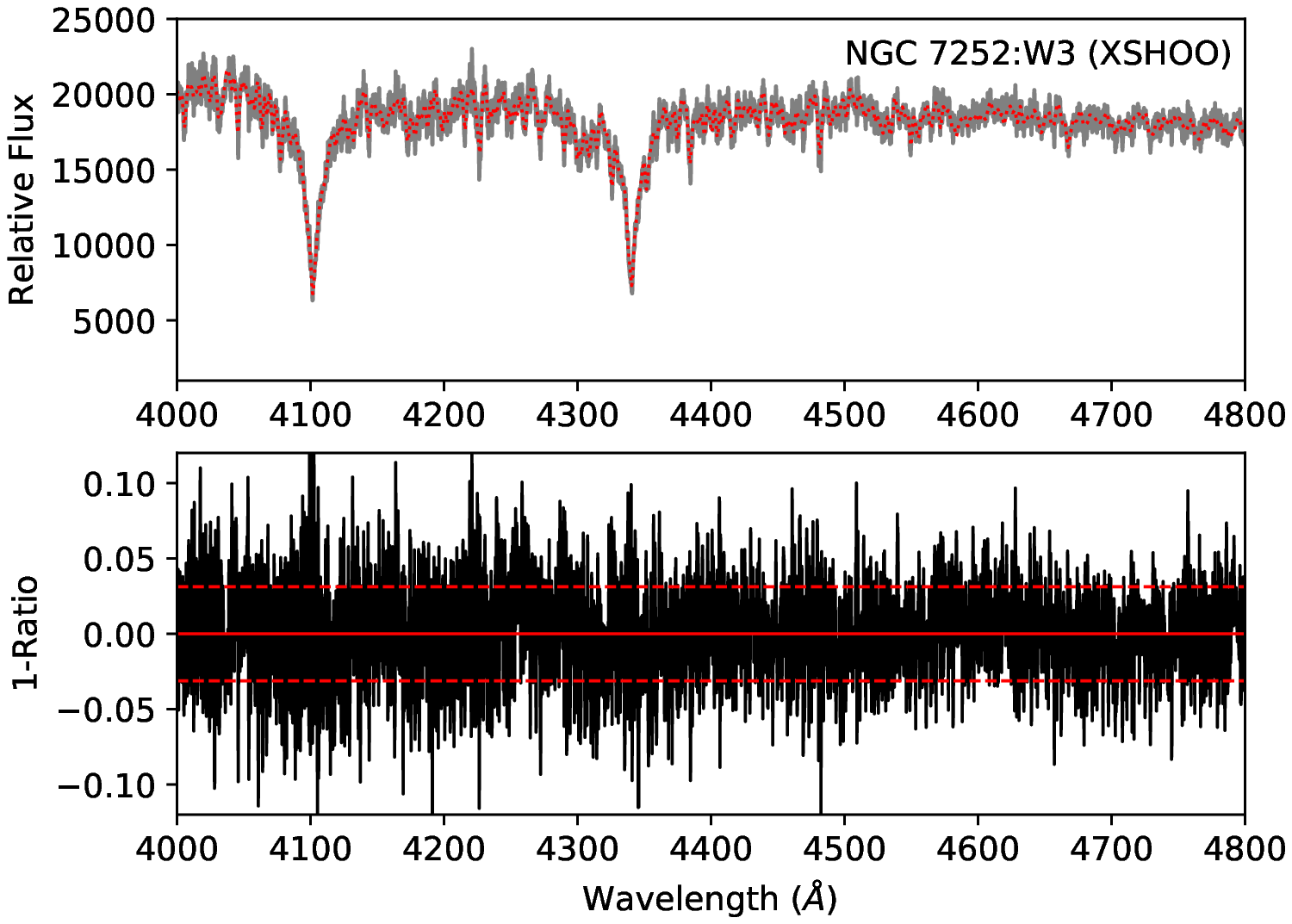}

\caption{The same as Fig.~\ref{fig:w3_muse} but now for the blue portion of the XShooter spectrum of W3.}
\label{fig:w3_xshoo}
\end{figure}

\subsection{The Effect of Dredge-Up}
\label{sec:dredge_up}

Due to the young age of W3 ($\sim600$~Myr - see next section) the surface abundances of some elements are expected to be affected by the first dredge-up of material from the stellar core to the surface after the star leaves the main-sequence.    In stellar models with masses above $\sim$1.5-2.0 $M_{\odot}$, the first dredge up brings into the convective envelope matter from inner layers involved in the Ne-Na cycle during the main sequence. In these layers the Na abundance (that has attained the equilibrium value of the Ne-Na cycle -  see, e.g., Denisenkov \& Denisenkova~1990) has increased compared to the initial value,  at the expenses of the much more abundant Ne. The net effect of the dredge up is then an increase of the surface Na during the RGB phase. This has been empirically observed in a sample of Galactic open clusters  (see, e.g. Smiljanic et al.~2016). For an age of 600~Myr and solar metallicity (close to the Hyades age and metal content) Smiljanic et al.~(2016) find an increase of Na after the dredge up by $\sim$0.2~dex.

To take this into account we also ran models that did not adopt the initial abundances, but rather the modifie surface abundance which includes the effects of the dredge-up.  For this we adopted an enhanced [Na/Fe] value of $+0.25$ for all He burning (and post He-burning) stars in the model.  This value was taken from Smiljanic et al.~(2016) based on the Na-enhancement seen in the Hyades open cluster of similar age and metallicity.

For populations with enhanced Na at birth, the surface increase of Na due the dredge up will be smaller (or reduced to zero)  than this value, because of a reduced difference between the Na equilibrium abundances, and the initial abundance in the deepening convective envelope.  We will explicitly mention when our modelling includes the first dredge-up throughout the paper.

\section{Results}
\label{sec:results}

\subsection{Searching for Na enhancement}
In Fig.~\ref{fig:nad} we show the spectral region of NaD for W3, along with three models (without dredge-up) for solar scaled (red), intermediate (green, [Na/Fe]$=+0.5$), and extreme (blue, [Na/Fe]$=+1.0$). From this we would infer that W3 is enhanced, albeit at a low level, in [Na/Fe], on average, with a value of [Na/Fe]$=0.2-0.3$~dex.  

\begin{figure}
\centering
\includegraphics[width=8cm]{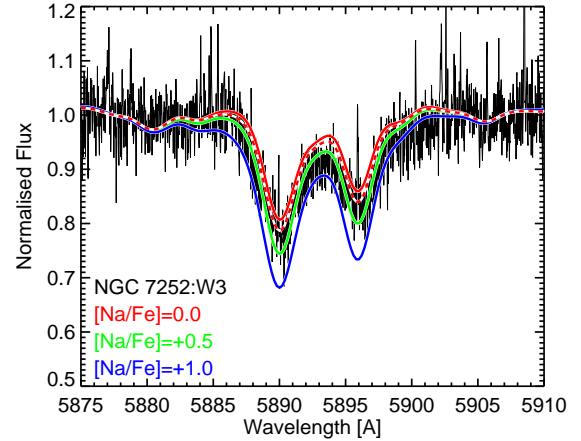}
\caption{The same as Fig.~\ref{fig:nad} but now the solar-scaled model that includes the effects of dredge-up is shown as a red-dashed line.  }
\label{fig:nad_dredge}
\end{figure}

However, as discussed in \S~\ref{sec:dredge_up}, at the age and metallicity of W3, stars leaving the main sequence are expected to undergo significant amounts of internal mixing, which will bring up processed material to the surface of the star.  At this age, this material is expected to be enhanced in [Na/Fe].  In Fig.~\ref{fig:nad_dredge} we show the effects of including the enhanced surface abundance of [Na/Fe] due to the dredge-up.  The intermediate and extreme models do not show significant changes to the strength of the NaD lines.  Effectively, the dredge-up causes the integrated light of the cluster to appear enhanced in [Na/Fe], by about $0.1-0.2$~dex.  This brings the best fitting {\sc Chrusher} model spectrum into agreement with the observations in the NaD line.


\subsection{Comparing W3 and W30}

\begin{figure}
\centering  
\includegraphics[width=8cm]{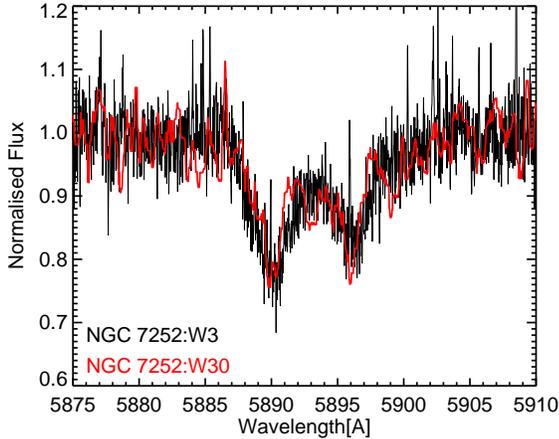}
\caption{A comparison between the NaD profile of W3 (black) and W30 (red - smoothed).  The spectrum of W30 has been boxcar smoothed with a width of seven pixels to decrease the noise.  Note that the two spectra are nearly identical, although the line width are slightly larger in W3 due to the higher velocity dispersion.  }
\label{fig:nad_both}
\end{figure}

While the S/N of the W30 spectrum is significantly lower than W3, we can compare the two directly, given their similar age and metallicity.  This can be inferred from the similarity of their optical/UV colours and optical low resolution spectra (Schweizer \& Seitzer~1998; Bastian et al.~2013), hence we can directly compare the line strengths/profiles of the two clusters. In Fig.~\ref{fig:nad_both} we show the integrated spectrum around the NaD lines for each cluster, with W3 in black and W30 in red.  As can be seen, the two spectra are quite similar, with the wider width of W3 expected due to the higher velocity dispersion (45~km/s compared to 27.5~km/s).  We can then infer, as was the case for W3, that W30 does not host a significant population of stars enhanced in [Na/Fe].

Note that if MPs were present in W3 and W30,  and assuming that the mass dependence of [Na/Fe] seen among Galactic GCs extends to the mass of W3, we would expect significantly weaker NaD lines in W30 than W3, due to the difference in mass (with W3 $\sim4\times$ more massive than W30) as seen in GCs in M31 (Schiavon et al.~2013; Sakari et al.~2016) and the Galaxy (Carretta et al.~2010; Milone et al.~2017).  The similarity in the lines corroborates our conclusion that MPs are not present in either cluster (at least at a detectable level).

\section{Discussion and Conclusions}
\label{sec:conclusions}

Recent work using resolved photometry and spectroscopy has shown that multiple populations are not restricted to ancient globular clusters but are present in clusters with ages down to (at least) $\sim2$~Gyr (Niederhofer et al. 2017; Hollyhead et al. 2017; 2019; Martocchia et al. 2018a; 2019; Li \& de Grijs 2019).  However, despite significant efforts on resolved photometric (Martocchia et al. 2018a) and spectroscopic surveys (Mucciarelli et al. 2008, 2011, 2014) no abundance spreads have been found in massive clusters with ages below $\sim2$~Gyr.  Additionally, integrated light studies of clusters with masses $\sim10^6$~\msun\ with ages of $30-100$~Myr whose spectra are dominated by red supergiants, also have not found evidence of MPs (Cabrera-Ziri et al.~2016b, Lardo et al.~2017).   For the resolved clusters that do show MPs (with ages $\gtrsim2$~Gyr), Martocchia et al. (2018b; 2019) note a trend with age in their sample of young, intermediate and ancient clusters in the LMC and SMC, with younger clusters showing smaller abundance spreads (and potentially smaller enriched fractions) than older clusters.  

The origin of this trend with age and the apparent step-change at $\sim2$~Gyr is currently unknown.  We have investigated the role that age plays in the onset of MPs by looking at three extremely massive clusters ($M >10^7$~\msun) that span this critical age.  In particular, in our sample we have two 
clusters ages $\sim600$~Myr (NGC~7252:W3 and W30) and another with an age of $\sim3$~Gyr (NGC~1316:G114; Bastian et al.~2019).  For the latter cluster $(\sim10^7$~\msun) we found an integrated abundance of [Na/Fe]$=0.6\pm0.2$, while for the similar mass (or higher) $\sim600$~Myr clusters we find [Na/Fe]$<0.2$.  We note that high resolution spectroscopic studies of metal rich (near solar) stars in the Galactic Bulge show variations in the [Na/Fe] abundances of 0.1 to 0.2 dex, which sets the lower limit to this technique, due to galactic chemical evolution and spatial variations in the enrichment of the ISM (e.g., Bensby et al.~2017).

By finding evidence of MPs (through enhanced [Na/Fe]) in the older cluster and not the two young clusters, our results support the findings of Martocchia et al.~(2018a; 2019) where the onset of MPs (or at least the ability to observe them) depends strongly on age, consistent with the previously suggested $\sim2$~Gyr limit.

These results demonstrate that the trend of MP properties with cluster age is not limited to the Magellanic Clouds, but rather that it is a seemingly general property of the phenomenon.  Much of the previous photometric work has adopted techniques that are mainly sensitive to spreads in [N/Fe].  While nitrogen is an element that is seen to show strong variations within Galactic globular clusters (e.g., Sneden et al.~1992), the nitrogen surface abundance is severely affected by the first dredge-up, a process that is age dependent and likely affects ongoing photometric surveys (e.g., Salaris et al.~2020).  By searching in [Na/Fe], the surface abundance of which is much less affected by the dredge-up, we have shown that the age dependence is physical and not simply due to internal stellar mixing.

\section*{Acknowledgments}

NB gratefully acknowledges financial support from the Royal Society (University Research Fellowship).  NB, CU, and SK gratefully acknowledge financial support from the European Research Council (ERC-CoG-646928, Multi-Pop).  CL acknowledges financial support from the Swiss National Science Foundation (Ambizione grant PZ00P2\_168065).  Support for this work was provided by NASA through Hubble Fellowship grant HST-HF2-51387.001-A awarded by the Space Telescope Science Institute, which is operated by the Association of Universities for Research in Astronomy, Inc., for NASA, under contract NAS5-26555. WC acknowledges funding from the Swiss National Science Foundation under grant P400P2\_183846.

\bsp
\label{lastpage}

\vspace{-0.5cm}
\begin{thebibliography}{99}


\bibitem[Bastian et al.(2006)]{2006A&A...448..881B} Bastian, N., Saglia, R.~P., Goudfrooij, P., et al.\ 2006, A\&A, 448, 881 

\bibitem[\protect\citeauthoryear{Bastian, et al.}{2013}]{2013MNRAS.431.1252B} Bastian N., Schweizer F., Goudfrooij P., Larsen S.~S., Kissler-Patig M., 2013, MNRAS, 431, 1252


\bibitem[Bastian \& Lardo(2018)]{2018ARA&A..56...83B} Bastian, N., \& Lardo, C.\ 2018, ARA\&A, 56, 83 

\bibitem[Bastian et al.(2019)]{2019MNRAS.489L..80B} Bastian, N., Usher, C., Kamann, S., et al.\ 2019, MNRAS, 489, L80

\bibitem[Bertelli et al.(1994)]{1994A&AS..106..275B} Bertelli, G., Bressan, A., Chiosi, C., Fagotto, F., \& Nasi, E.\ 1994, A\&AS, 106, 275 

\bibitem[Cabrera-Ziri et al.(2016)]{2016MNRAS.457..809C} Cabrera-Ziri, I., Bastian, N., Hilker, M., et al.\ 2016a, MNRAS, 457, 809 

\bibitem[Cabrera-Ziri et al.(2016)]{2016MNRAS.460.1869C} Cabrera-Ziri, I., Lardo, C., Davies, B., et al.\ 2016b, MNRAS, 460, 1869 

\bibitem[Carretta et al.(2010)]{2010A&A...516A..55C} Carretta, E., Bragaglia, A., Gratton, R.~G., et al.\ 2010, A\&A, 516, A55 






\bibitem[Conroy \& Gunn(2010)]{2010ApJ...712..833C} Conroy, C., \& Gunn, J.~E.\ 2010, ApJ, 712, 833 



\bibitem[\protect\citeauthoryear{Denisenkov \& Denisenkova}{1990}]{1990SvAL...16..275D} Denisenkov P.~A., Denisenkova S.~N., 1990, SvAL, 16, 275


\bibitem[Goudfrooij et al.(2001)]{2001MNRAS.322..643G} Goudfrooij, P., Mack, J., Kissler-Patig, M., Meylan, G., \& Minniti, D.\ 2001, MNRAS, 322, 643 



\bibitem[Hollyhead et al.(2017)]{2017MNRAS.465L..39H} Hollyhead, K., Kacharov, N., Lardo, C., et al.\ 2017, MNRAS, 465, L39 

\bibitem[Hollyhead et al.(2019)]{2019MNRAS.484.4718H} Hollyhead, K., Martocchia, S., Lardo, C., et al.\ 2019, MNRAS, 484, 4718 


\bibitem[Koleva et al.(2008)]{2008MNRAS.385.1998K} Koleva, M., Prugniel, P., Ocvirk, P., Le Borgne, D., \& Soubiran, C.\ 2008, MNRAS, 385, 1998 

\bibitem[Kroupa(2001)]{2001MNRAS.322..231K} Kroupa, P.\ 2001, MNRAS, 322, 231 





\bibitem[Lardo et al.(2017)]{2017MNRAS.468.2482L} Lardo, C., Cabrera-Ziri, I., Davies, B., \& Bastian, N.\ 2017, MNRAS, 468, 2482 




\bibitem[Le Borgne et al.(2004)]{2004A&A...425..881L} Le Borgne, D., Rocca-Volmerange, B., Prugniel, P., et al.\ 2004, A\&A, 425, 881 

\bibitem[\protect\citeauthoryear{Li \& de Grijs}{2019}]{2019ApJ...876...94L} Li C., de Grijs R., 2019, ApJ, 876, 94


\bibitem[Maraston et al.(2004)]{2004A&A...416..467M} Maraston, C., Bastian, N., Saglia, R.~P., et al.\ 2004, A\&A, 416, 467 

\bibitem[Martocchia et al.(2017)]{2017MNRAS.468.3150M} Martocchia, S., Bastian, N., Usher, C., et al.\ 2017, MNRAS, 468, 3150 

\bibitem[Martocchia et al.(2018)]{2018MNRAS.473.2688M} Martocchia, S., Cabrera-Ziri, I., Lardo, C., et al.\ 2018a, MNRAS, 473, 2688

\bibitem[\protect\citeauthoryear{Martocchia, et al.}{2018}]{2018MNRAS.477.4696M} Martocchia S., et al., 2018b, MNRAS, 477, 4696

\bibitem[\protect\citeauthoryear{Martocchia, et al.}{2019}]{2019MNRAS.487.5324M} Martocchia S., et al., 2019, MNRAS, 487, 5324

\bibitem[Miller et al.(1997)]{1997AJ....114.2381M} Miller, B.~W., Whitmore, B.~C., Schweizer, F., \& Fall, S.~M.\ 1997, AJ, 114, 2381 

\bibitem[Milone et al.(2017)]{2017MNRAS.464.3636M} Milone, A.~P., Piotto, G., Renzini, A., et al.\ 2017, MNRAS, 464, 3636 

\bibitem[\protect\citeauthoryear{Mucciarelli, et al.}{2008}]{2008AJ....136..375M} Mucciarelli A., Carretta E., Origlia L., Ferraro F.~R., 2008, AJ, 136, 375

\bibitem[\protect\citeauthoryear{Mucciarelli, et al.}{2011}]{2011MNRAS.413..837M} Mucciarelli A., et al., 2011, MNRAS, 413, 837

\bibitem[\protect\citeauthoryear{Mucciarelli, et al.}{2014}]{2014ApJ...793L...6M} Mucciarelli A., Dalessandro E., Ferraro F.~R., Origlia L., Lanzoni B., 2014, ApJL, 793, L6


\bibitem[Niederhofer et al.(2017)]{2017MNRAS.465.4159N} Niederhofer, F., Bastian, N., Kozhurina-Platais, V., et al.\ 2017, MNRAS, 465, 4159 


\bibitem[Prugniel \& Soubiran(2001)]{2001A&A...369.1048P} Prugniel, P., \& Soubiran, C.\ 2001, A\&A, 369, 1048 





\bibitem[\protect\citeauthoryear{Sakari, et al.}{2016}]{2016ApJ...829..116S} Sakari C.~M., et al., 2016, ApJ, 829, 116

\bibitem[Salaris et al.(2020)]{2020MNRAS.492.3459S} Salaris, M., Usher, C., Martocchia, S., et al.\ 2020, MNRAS, 492, 3459

\bibitem[\protect\citeauthoryear{Schiavon, et al.}{2013}]{2013ApJ...776L...7S} Schiavon R.~P., et al., 2013, ApJL, 776, L7

\bibitem[Schweizer \& Seitzer(1998)]{1998AJ....116.2206S} Schweizer, F., \& Seitzer, P.\ 1998, AJ, 116, 2206 

\bibitem[\protect\citeauthoryear{Smiljanic, et al.}{2016}]{2016A&A...589A.115S} Smiljanic R., et al., 2016, A\&A, 589, A115

\bibitem[]{} Sneden C, Kraft RP, Prosser CF, Langer GE. 1992., AJ, 104, 2121






\end{thebibliography}
\end{document}